\documentclass[12pt]{iopart}

\expandafter\let\csname equation*\endcsname\relax
\expandafter\let\csname endequation*\endcsname\relax
\usepackage{amsmath}  

\usepackage{nicefrac}
\usepackage{lineno} 

\usepackage[usenames, dvipsnames]{xcolor}
\usepackage{subcaption}

\usepackage{tikz}
\usepackage{tikz-3dplot}
\usepackage[font=small,labelfont=bf]{caption}\usepackage{subcaption}
\usepackage{graphicx}
\usepackage{cite}

\usepackage[]{hyperref}
\hypersetup{
    colorlinks,%
    citecolor=blue,%
    filecolor=black,%
    linkcolor=red,%
    urlcolor=black
}

\usepackage{cleveref}
\usepackage{tabularx}
\usepackage{multicol} 
\usepackage[bold]{hhtensor}
\usetikzlibrary{arrows,decorations.pathmorphing,backgrounds,positioning,fit,matrix}
\usetikzlibrary{intersections,calc}
\usetikzlibrary{shapes.misc}
\usepackage{pst-eucl}
\tikzset{cross/.style={cross out, draw=black, fill=none, minimum size=2*(#1-\pgflinewidth), inner sep=0pt, outer sep=0pt}, cross/.default={2pt}}

\begin{document}

\title[Zhao and Marian]{Direct prediction of the solute softening-to-hardening transition in W-Re alloys using stochastic simulations of screw dislocation motion}

\author{Yue Zhao$^1$ and Jaime Marian$^{1,2}$}
\address{$^1$ Department of Materials Science and Engineering, University of California Los Angeles, Los Angeles, CA 90095}
\address{$^2$ Department of Mechanical and Aerospace Engineering, University of California Los Angeles, Los Angeles, CA 90095}
\ead{jmarian@ucla.edu}
\vspace{10pt}
\begin{indented}
\item[]\today
\end{indented}

\begin{abstract}
Interactions among dislocations and solute atoms are the basis of several important processes in metals plasticity.
In body-centered cubic (bcc) metals and alloys, low-temperature plastic flow is controlled by screw dislocation glide, which is known to take place by the nucleation and sideward relaxation of kink pairs across two consecutive \emph{Peierls} valleys. In alloys, dislocations and solutes affect each other's kinetics via long-range stress field coupling and short-range inelastic interactions. It is known that in certain substitutional bcc alloys a transition from solute softening to solute hardening is observed at a critical concentration. In this paper, we develop a kinetic Monte Carlo model of screw dislocation glide and solute diffusion in substitutional W-Re alloys. We find that dislocation kinetics is governed by two competing mechanisms. At low solute concentrations, nucleation is enhanced by the softening of the Peierls stress, which overcomes the elastic repulsion of Re atoms on kinks. This trend is reversed at higher concentrations, resulting in a minimum in the flow stress that is concentration and temperature dependent. This minimum marks the transition from solute softening to hardening, which is found to be in reasonable agreement with experiments. 
\end{abstract}

%
%
%
%
%

\section{Introduction}
\label{intro}

The applications of tungsten (W) and its alloys have been limited to a great extent by their brittleness at ambient temperatures. Typical ductile-to-brittle transition temperatures (DBTT) for arc-melted W range from approximately 475 to 525 K, whereas for wrought and recrystallized metal the DBTT falls between 560 to 645 K \cite{raffo1969yielding,Stephens1970}. Some authors have measured a DBTT in excess of 1000 K for sintered specimens \cite{davis1998assessment,faleschini2007fracture}.
At room temperature, the fracture toughness $K_c$ of polycrystalline W in ranges between 5.1 (sintered) and 9.1 (rolled) MPa$\cdot$$\sqrt{\rm m}$ \cite{Gludovatz2010674}. 
W is typically alloyed with 5-26 at.\% Re to increase low temperature ductility and improve high temperature strength and plasticity, which can increase $K_c$ by up to one order of magnitude \cite{Mutoh1995,Faleschini2007800}

The physical origins behind the Re-induced ductilization have been discussed in the literature \cite{Stephens1970,LUO1991107,romaner2010effect,wurster2010high} and point in some way or another to alterations in the core structure of $\nicefrac{1}{2}\langle111\rangle$ screw dislocations, which both reduce the effective Peierls stress and extend the number of possible slip pathways. However, this is a highly local effect, concentrated around the vicinity of screw dislocation cores, whereas plasticity involves both long-range and local solute-core interactions. Although the softening of the Peierls stress, $\sigma_P$, can be expected to lead to an increased screw dislocation mobility, factors such as kink-pair nucleation enthalpies and kink velocities must be considered to accurately assess whether Re additions result in global solute softening, hardening, or both. 
This was recognized early on for body-centered cubic (bcc) alloys in the experimental literature, where there is ample evidence of a transition from solute softening to solute strengthening in substitutional alloys \cite{RAVI1969547,RAVI1970623,Leslie1972,GIBALA19731143,PINK19801,LUO1991107,Okazaki1996}. This transition is usually rationalized as a competition between kink-pair nucleation and kink motion, which are thought to depend inversely on solute content. Although early models built on this idea to successfully predict the softening-to-hardening transition \cite{SATO1973753,KIRCHHEIM2012767}, it was not until very recently that sophisticated electronic structure calculations have revealed the local effect of Re alloying on the structure of screw dislocation cores in W \cite{romaner2010effect,wurster2010high,0953-8984-25-2-025403,PhysRevB.95.094114}.  The effect on solute migration of dislocation stress fields can now also be characterized by combining elasticity theory with electronic structure calculations \cite{YASI20105704,hossain2014stress,1212175,HU2017304}.  

The purpose of this study is to predict the softening-to-hardening transition in W-Re alloys by directly simulating the interaction of screw dislocations with solute atoms. We develop a kinetic Monte Carlo (kMC) model that accounts for thermally-activated kink-pair nucleation and solute diffusion via stress-field coupling and short-range inelastic interactions. Within our model the solute subsystem and the dislocation evolve in a coupled fashion, with solute diffusion being sensitive to dislocation stress fields, and kink-pair nucleation and propagation being influenced by the presence of solute. Our model is parameterized entirely using electronic structure calculations, as described by Hossain and Marian \cite{hossain2014stress}, and is significantly more efficient than direct atomistic simulations. This allows us to study the relevant parameter space of stress, temperature, Re content, and dislocation line length by simulating the motion of a screw dislocation in a random W-Re solid solution. 
In Section \ref{methods}, we provide a description of the methods employed, followed by the results of dislocation mobility and critical stresses in Section \ref{results}. We conclude in Section \ref{disc} with a discussion of our findings and the conclusions.

\section{Numerical model and parameterization}
\label{methods}
Detailed descriptions of the kMC model and the electronic structure calculations of model parameters have been published previously \cite{hossain2014stress,stukowski2015thermally}, and here we simply provide a brief review and highlight the aspects that are most relevant to the coupling between the dislocation and the solute subsystem.  

\subsection{Effect of solute on kink-pair activation enthalpy}
We start with the expression for the nucleation rate of a kink pair on a screw dislocation segment in the pure material:
\begin{equation}
r_{\rm kp}=\nu_0\frac{L-w}{b}\exp\left(-\frac{\Delta H(\matr{\sigma})}{kT}\right)
\label{kprate}
\end{equation}
where $\nu_0$ is an attempt frequency, $L$ is the available segment length, $w$ is the kink pair separation, $b=a\sqrt{3}/2$ is the modulus of the Burgers vector $\vec{b}$ ($a$ is the lattice parameter), $\Delta H$ is the activation enthalpy, which is a function of the local stress tensor $\matr{\sigma}$, $k$ is Boltzmann's constant and $T$ is the absolute temperature. For its part, the kink-pair activation enthalpy is assumed to follow the expression:
\begin{equation}
\Delta H(\matr{\sigma})=\Delta H_0\left(1-\left[\Theta(\matr{\sigma})\right]^p\right)^q
\label{deltah}
\end{equation}
where is $\Delta H_0$ the sum of the formation enthalpy of two opposite-signed kinks, $\Theta(\matr{\sigma})$ is a \emph{stress ratio} that depends on $\matr{\sigma}$, and $p$ and $q$ are fitting parameters. The stress ratio is defined as:
\begin{equation}
\Theta(\matr{\sigma})=\frac{\tau_{\rm MRSS}}{\tau_c(\matr{\sigma})}
\label{thetaT}
\end{equation}
where $\tau_{\rm MRSS}$ is the \emph{maximum} resolved shear stress (MRSS) and $\tau_c(\matr{\sigma})$ is the critical stress for athermal dislocation motion. The stress factor includes in its formulation all non-Schmid effects characteristic of bcc metals, given below.

Equation \eref{deltah} is modified by the presence of solute. Specifically, when a Re atom is found adjacent to the dislocation line, an extra (binding) energy has to be overcome to nucleate the next kink-pair along that dislocation segment. This is shown in Figure \ref{Re1}, where a Re atom is along the path of the dislocation. In such case, eq. \eref{deltah} is modified as:
\begin{figure}[h]
	\centering
\includegraphics[width=0.85\columnwidth]{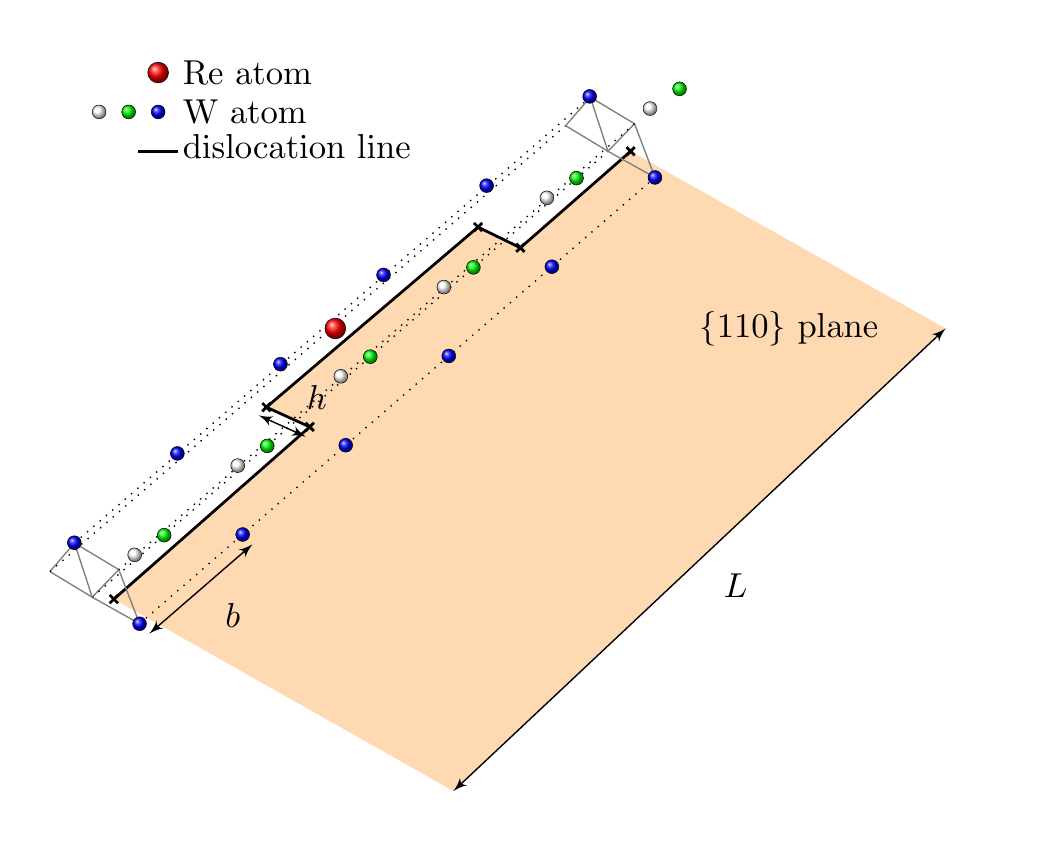}
\caption{Position of lattice atoms relative to a screw dislocation segment of length $L$ containing a kink-pair. The presence of a Re atom (red sphere) changes the activation enthalpy of subsequent kink-pair nucleation events. W atoms are colored according to  the $\langle111\rangle$ string they belong to.}
	\label{Re1}
\end{figure}
\begin{equation}
\Delta H'(\matr{\sigma})=e_b^{\rm kp}+\Delta H(\matr{\sigma})
\label{deltahprime}
\end{equation}
The binding energy $e_b^{\rm kp}$ depends on the distance and position of the solute atom relative to the dislocation core, and we have calculated it using DFT calculations \cite{hossain2014stress}. It takes a value of 0.25 eV for the configuration shown in Fig.\ \ref{Re1} and decays quickly to zero thereafter, taking a value of 0.04 eV for the next nearest atomic row.

\subsection{Effect of solute on kink motion}

In the pure metal, kinks move along the dislocation line with a velocity:
\begin{equation}
v_{\rm k}=\frac{\matr{\sigma}\vec{b}}{B}
\label{visc}
\end{equation}
where $B$ is a friction coefficient. This reflects the fast viscous motion of nonscrew segments. After nucleation, a kink will take an average time of:
\begin{equation}
\delta t_{\rm k}=\frac{L-w}{2v_{\rm k}}=\frac{B(L-w)}{2\matr{\sigma}\vec{b}}
\label{tkp}
\end{equation}
to sweep half of the segment length of the dislocation.
However, when a moving kink encounters a Re atom, it binds to it with an energy that depends on their relative location. From that moment on, kink motion ceases and detachment becomes thermally activated. The corresponding detachment rate is:
\begin{equation}
r_{\rm sk}=\nu_1\exp\left(-\frac{e^{\rm sk}_b}{kT}\right)
\label{detrap}
\end{equation}
where $v_1$ is an attempt frequency (different from $\nu_0$) and $e_b^{\rm sk}$ is the binding energy between the kink segment and the Re atom. Kink segments have partial edge dislocation character and, as such, induce both tensile and compressive stresses. Re atoms are approximately 3\% larger than W atoms and therefore introduce a certain dilation strain. Therefore, Re binds preferentially to the tensile lobe of a single kink segment deformation field, with a value of $e^{\rm sk}_b=0.45$ eV. In any case, the binding energy in the compressive region is also attractive, with a maximum value of 0.35 eV \cite{hossain2014stress}. Thus, for simplicity, we take an average value of 0.40 eV regardless of solute atom relative location.

Once a dissociation event occurs for a kink-solute configuration, the kink resumes its viscous motion according to eq.\ \eref{visc}.

\subsection{The critical stress and non-Schmid effects}
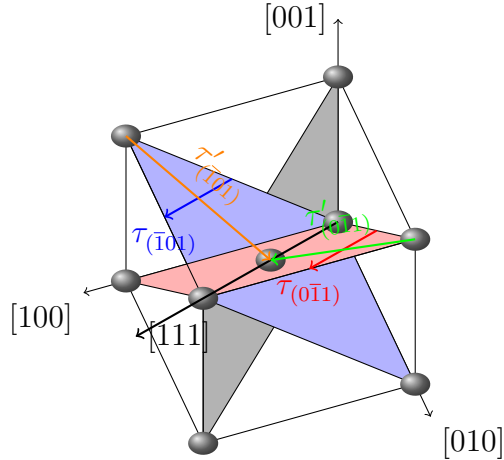
\begin{figure}[h]
	\centering
%
\tdplotsetmaincoords{40}{160}	
\begin{tikzpicture}[scale=1,tdplot_main_coords]
\def \L {3}
\def \R{0.2}
\coordinate (O) at (0,0,0);
\draw[->] (O) -- (1.2*\L,0,0) node[anchor=north east]{$[100]$};
\draw[->] (O) -- (0,1.2*\L,0) node[anchor=north west]{$[010]$};
\draw[->] (O) -- (0,0,1.4*\L) node[anchor=east]{$[001]$};
\tikzstyle{bluefill} = [fill=black!30,fill opacity=0.3]
\tikzstyle{greenfill} = [fill=blue!30,fill opacity=0.3]
\tikzstyle{redfill} = [fill=red!30,fill opacity=0.3]
\filldraw[bluefill] (0,0,0) -- (\L,\L,0) -- (\L,\L,\L) -- (0,0,\L) -- cycle;
\filldraw[greenfill] (0,0,0) -- (0,\L,0) -- (\L,\L,\L) -- (\L,0,\L) -- cycle;
\filldraw[redfill] (0,0,0) -- (\L,0,0) -- (\L,\L,\L) -- (0,\L,\L) -- cycle;
\draw[] (0,0,\L) -- (\L,0,\L);
\draw[] (\L,0,0) -- (\L,\L,0);
\draw[] (\L,0,\L) -- (\L,0,0);
\draw[] (\L,\L,0) -- (\L,\L,\L);
\draw[] (\L,\L,\L) -- (\L,0,\L);
\draw[] (\L,\L,\L) -- (0,\L,\L);
\draw[] (\L,\L,0) -- (0,\L,0);
\draw[] (0,\L,0) -- (0,\L,\L);
\draw[] (0,\L,\L) -- (0,0,\L);
\shade[ball color=gray] (0,0,0) circle(\R);
\shade[ball color=gray] (\L,0,0) circle(\R);
\shade[ball color=gray] (\L,\L,0) circle(\R);
\shade[ball color=gray] (0,\L,0) circle(\R);
\shade[ball color=gray] (0,0,\L) circle(\R);
\shade[ball color=gray] (\L,0,\L) circle(\R);
\shade[ball color=gray] (\L,\L,\L) circle(\R);
\shade[ball color=gray] (0,\L,\L) circle(\R);
\shade[ball color=gray] (0.5*\L,0.5*\L,0.5*\L) circle(\R);
\draw[thick,->] (0,0,0) -- (3*\L/2,3*\L/2,3*\L/2) node[above,right,sloped]{$[111]$};
\draw[thick,blue,->] (\L/2,0,\L/2) -- (\L,\L/2,\L) node[below,sloped] {$\tau_{(\overline{1}01)}$};
\draw[thick,orange,->] (\L,0,\L) -- (\L/2,\L/2,\L/2)node[above,midway,sloped] {$\tau_{(\overline{1}01)}'$};
\draw[thick,red,->] (0,\L/2,\L/2) -- (\L/2,\L,\L) node[below,sloped] {$\tau_{(0\overline{1}1)}$};
\draw[thick,green,->] (0,\L,\L) -- (\L/2,\L/2,\L/2) node[above,midway,sloped] {$\tau_{(0\overline{1}1)}'$};
\end{tikzpicture}
%
\caption{Stress projections as defined in eq.\ \eref{array} within the bcc unit cell. The dislocation line is assumed to be along the $[111]$ direction.}
	\label{coords}
\end{figure}

Our model includes non-Schmid effects from both sources commonly considered in the literature: the \emph{twinning/antitwinning} effect and non-glide stresses \cite{groger2008multiscale}. We follow our previous work on the topic \cite{cereceda2016unraveling,po2016phenomenological} to define the stress ratio $\Theta$ as: 
\begin{equation}
\Theta(\matr{\sigma})=\frac{
\tau_{\left(\bar{1}01\right)}+a_1\tau_{\left(0\bar{1}1\right)}
}
{
a_0\sigma_P(c)f\left(
\frac{
a_2\tau'_{\left(\bar{1}01\right)}+a_3\tau'_{\left(0\bar{1}1\right)}
}
{
a_0\sigma_P(c)
}
\right)
}
\label{ratio}
\end{equation}
where
\begin{eqnarray}
\tau_{\left(\bar{1}01\right)}=\matr{\sigma}:\vec{m}^{\alpha}\otimes\vec{n}^{\alpha} \nonumber\\
\tau_{\left(0\bar{1}1\right)}=\matr{\sigma}:\vec{m}^{\alpha}\otimes\vec{n}_1^{\alpha} \nonumber\\
\tau'_{\left(\bar{1}01\right)}=\matr{\sigma}:(\vec{n}^\alpha\times\vec{m}^{\alpha})\otimes\vec{n}^{\alpha} \nonumber\\
\tau'_{\left(0\bar{1}1\right)}=\matr{\sigma}:(\vec{n}_1^\alpha\times\vec{m}^{\alpha})\otimes\vec{n}_1^{\alpha} \label{array}
\end{eqnarray}
and $f(x)=2/\left(1+e^{2x}\right)$ \cite{po2016phenomenological}. The vectors $\vec{m}^\alpha$, $\vec{n}^{\alpha}$, and $\vec{n}_1^{\alpha}$ refer, respectively, to the slip direction, the glide plane, and the adjacent $\{110\}$ plane forming $+60^{\circ}$ about the screw direction to the glide plane. $\alpha$ refers to any one of the 12 independent slip systems in the bcc crystal lattice \cite{cereceda2016unraveling}. Figure \ref{coords} shows the relative geometry for the particular case of the $\nicefrac{1}{2}[111](\bar{1}01)$ slip system. We assume that the only dependence of $\Theta$ on the Re concentration $c$ is via an explicit dependence of the Peierls stress on $c$, which we discuss below.

\subsection{The Peierls stress}
First principles calculations in pure W put the value of the Peierls stress between 1.7 and 2.8 GPa \cite{romaner2010effect,samolyuk2012influence,dezerald2015first}. In this work, the key dependence of $\sigma_P$ is on the solute concentration, for which we use the calculations by Romaner \etal~\cite{romaner2010effect} and fit a numerical expression of the type $\sigma_P(c)=\frac{b_1}{c+b_2}$ (where $b_1$ and $b_2$ are fitting constants) to their data. 
This softening of the Peierls stress with Re content, together with a dislocation core transformation, has been used by the authors in ref.\ \cite{romaner2010effect} to explain the ductilization of W with Re in W-Re alloys.

\begin{figure}[h]
	\centering
	\includegraphics[width=0.7\columnwidth]{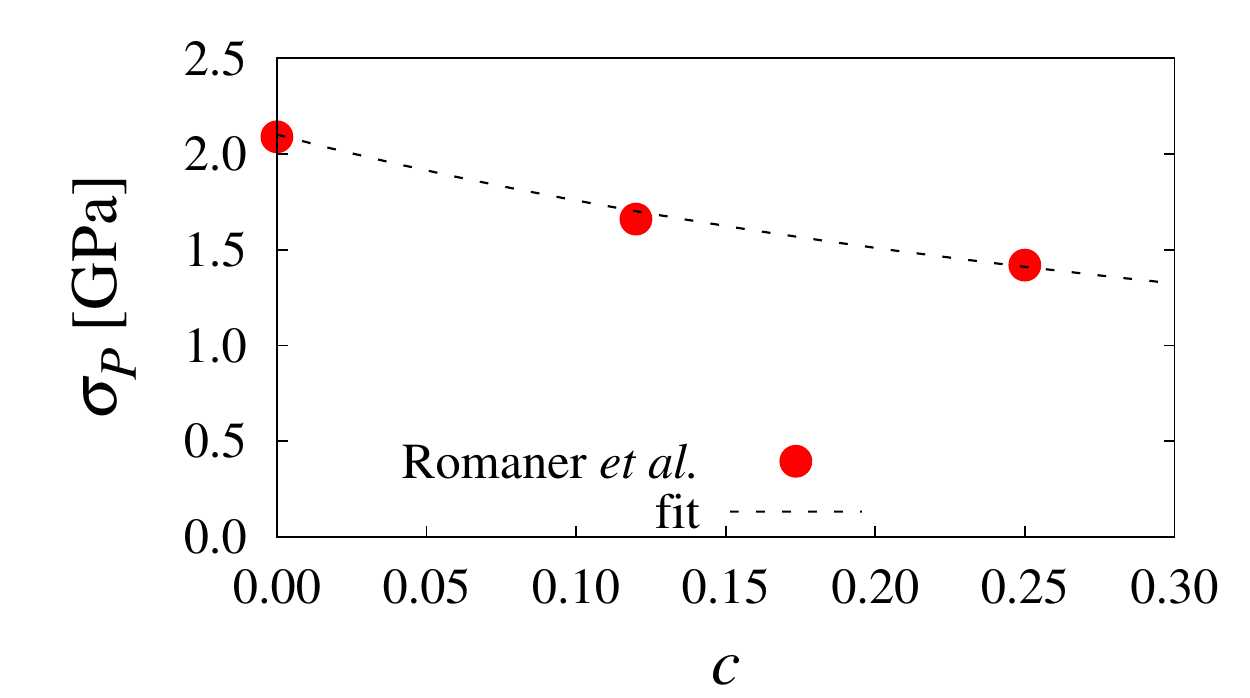}
	\caption{Variation of the Peierls stress with Re concentration from ref.\ \cite{romaner2010effect}. The data points have been fitted to the expression $\sigma_P(c)=\frac{1.07}{c+0.51}$ (in GPa).}
	\label{fig1}
\end{figure}

The fit, shown in Figure \ref{fig1}, yields values of $b_1=1.07$ (GPa) and $b_2=0.51$. This is the expression that we use in the remainder of this paper to carry out our kMC simulations in the presence of solute.

\subsection{Solute diffusion}

Re atoms occupying substitutional lattice sites in bcc W can diffuse by thermally activated nearest-neighbor jumps. This jump rate can be expressed as:
\begin{equation}
r_s=\nu_2\exp\left(-\frac{\Delta H_f^{\rm v}(\matr{\sigma})+H_m(\matr{\sigma})+e_b^{\alpha}}{kT}\right)
\label{diff}
\end{equation}
where $\nu_2$ is the attempt frequency for vacancy jumps --taken here as the \emph{Debye} frequency, cf.~Table \ref{tab-param}--, $\Delta H_f^{\rm v}$ is the vacancy formation enthalpy, $\Delta H_m$ is the migration enthalpy, both of which depend on the local stress tensor $\sigma$, and $e_b^{\alpha}$ is the binding energy either to a screw segment ($\alpha\equiv{\rm kp}$) or to a kink segment ($\alpha\equiv{\rm sk}$). This last term is only appropriate when the Re atom is bound to the dislocation core, otherwise it need not be included.  
The functional dependencies of the activation enthalpies on stress have been given in our previous publication \cite{hossain2014stress}, and they are based on the concept of formation or activation volume, which is a tensor defined as;
$$V^{*}_{ij}=\frac{\partial H}{\partial\sigma_{ij}}$$ 
such that the enthalpy is in general defined as:
$$\Delta H(\matr{\sigma})=E-\matr{\sigma}:\matr{V}^{*}$$
where the last term in the r.h.s~of the above expression is known as the \emph{mechanical work}. The vacancy formation and activation volumes in W-Re alloys have also been calculated in the dilute limit using DFT as given in ref.\ \cite{hossain2014stress}.
In the bcc lattice, each solute atom can jump to eight different nearest neighbor positions, each with its own rate as given by the local stress tensor at each location. To preserve detailed balance in the kMC simulations all eight possible transitions are considered for each Re atom.

The above approach results in tracer diffusion with a drift produced by the dislocation stress field, such that solute diffusion and dislocation motion are coupled to one another. This is the basic premise underlying a series of phenomena categorized under the umbrella of \emph{dynamic strain aging}, whereby coupled dislocation-solute evolution may result in anomalous mechanical behavior. 
\begin{table}
\centering
\caption{\label{tab-param}Parameter values employed in this work.}
\footnotesize
\begin{tabular}{@{}llc}
\br
Parameter & Units & Value \\
\mr
$a$ & \AA & 3.16 \\
$b$ & \AA & 2.74 \\
$h$ & \AA & 2.58 \\
$\nu_0$ & Hz & $9.1\times10^{11}$ \\
$\nu_1$ & Hz & $1.5\times10^{12}$ \\
$\nu_2$ & Hz & $6.4\times10^{12}$ \\
$w$ & $b$ & 11 \\
$\Delta H_0$ & eV & 1.63 \\
$p$ & - & 0.86 \\
$q$ & - & 1.69 \\
$B$ & Pa$\cdot$s & $8.3\times10^{-5}$ \\
$a_0$ & - & 1.45 \\
$a_1$ & - & 0.99 \\
$a_2$ & - & 2.36 \\
$a_3$ & - & 4.07 \\
$e_b^{\rm kp}$ & eV & 0.25 \\	
$e_b^{\rm sk}$ & eV & 0.40 \\	
\br
\end{tabular}\\
\end{table}
\normalsize
\begin{figure}[h]
	\centering
	\includegraphics[width=0.8\columnwidth]{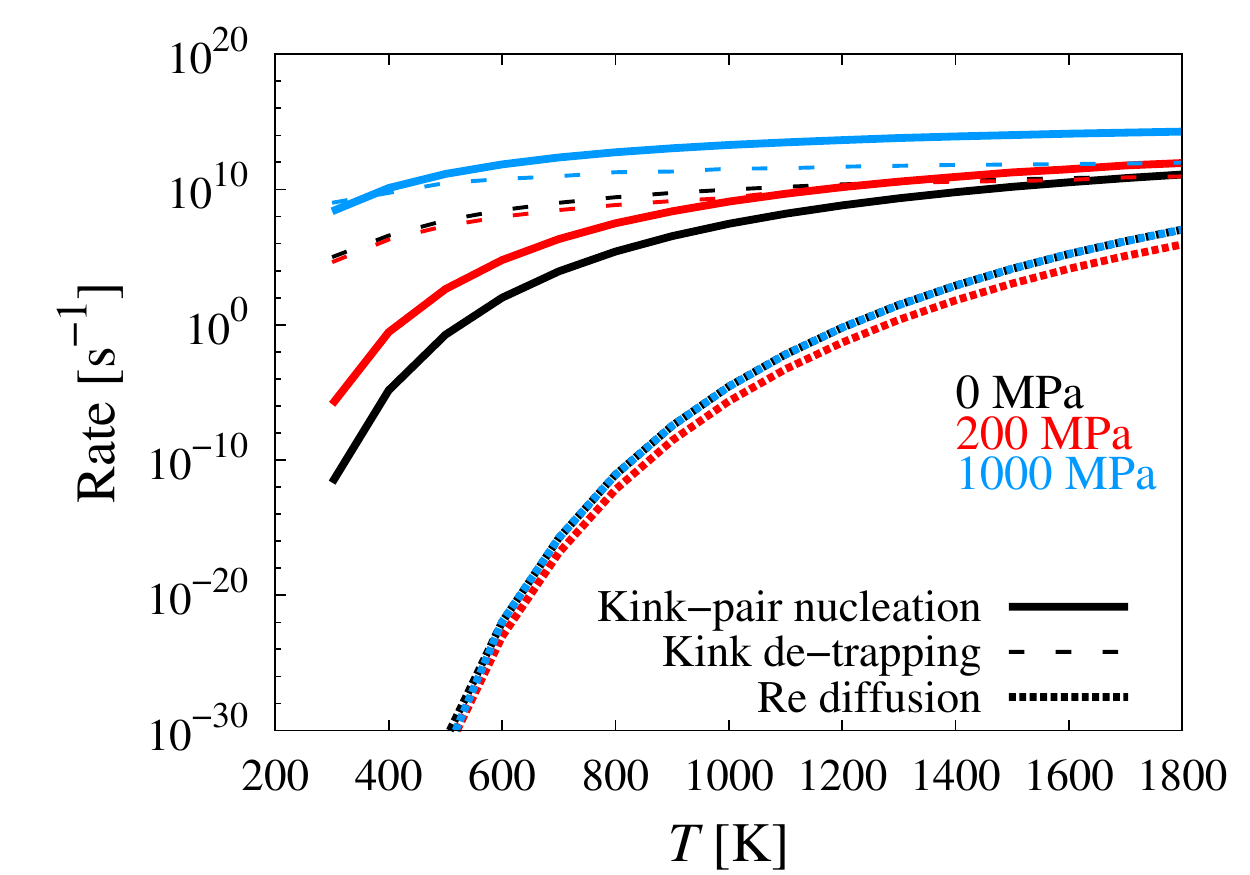}
	\caption{Magnitude of the three principal thermally-activated event rates as a function of temperature at three stress points. In the range explored, solute diffusivity is markedly lower than kink-pair nucleation and kink de-trapping.}
	\label{ratefig}
\end{figure}

However, a quick inspection of the three principal thermally-activated event rates, namely kink-pair nucleation, eq.~\eref{kprate}, kink de-trapping, eq.~\eref{detrap}, and solute hopping, eq.~\eref{diff} reveals a large gap in time scale between Re atom diffusion and the other two. This is showcased in Figure \ref{ratefig}, where the rates are plotted as a function of temperature at 0, 200, and 1000 MPa.
For this reason, in the following, we neglect solute diffusion and assume that the dislocations move in a static solute field.

\subsection{Solute distribution spatial update} 

\begin{figure*}[t!]
    \centering
    \begin{subfigure}[t]{0.25\textwidth}
        \centering
        \includegraphics[height=1.5in]{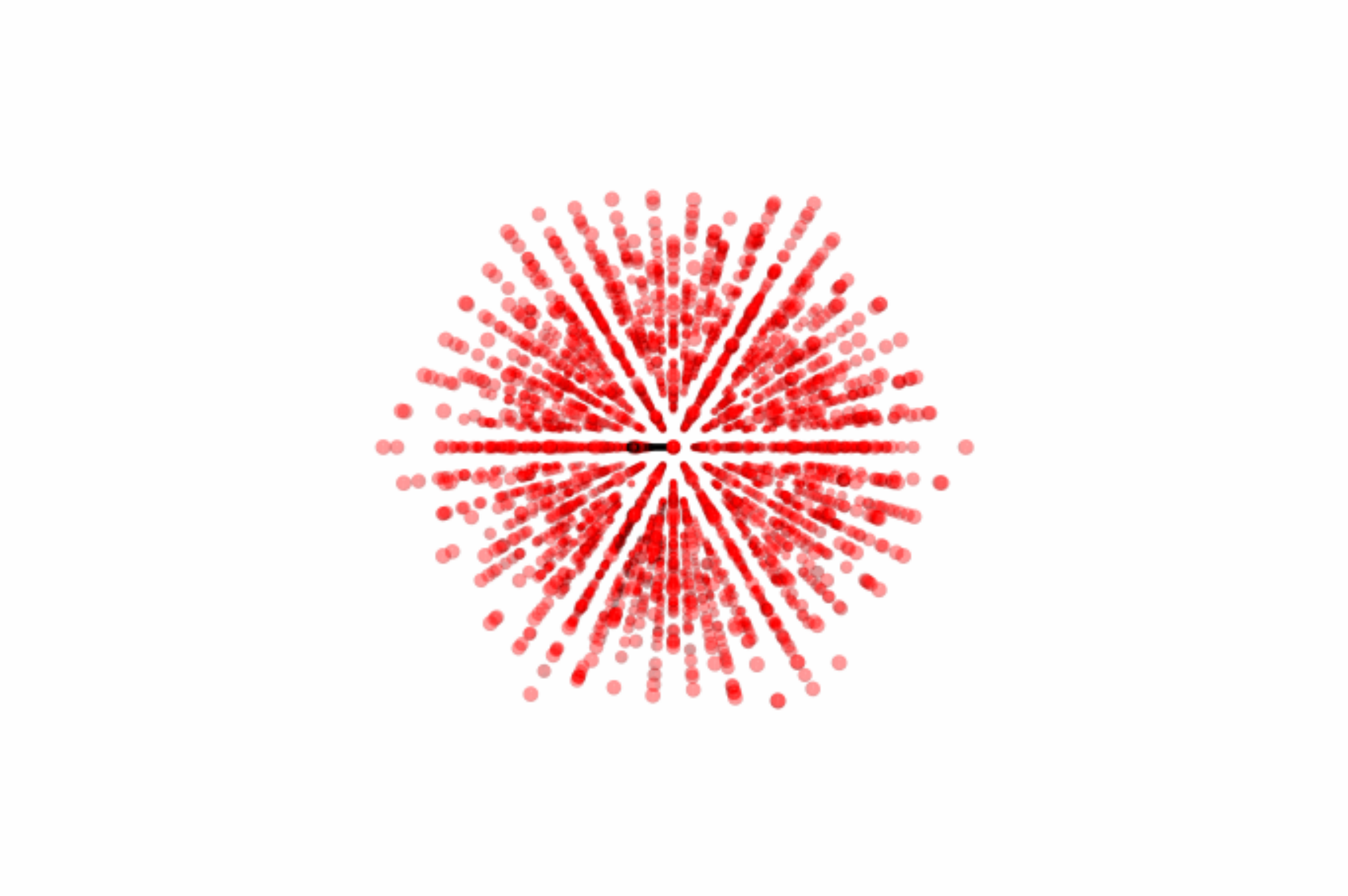}
        \caption{}
    \end{subfigure}
    \begin{subfigure}[t]{0.7\textwidth}
        \centering
        \includegraphics[height=2.9in]{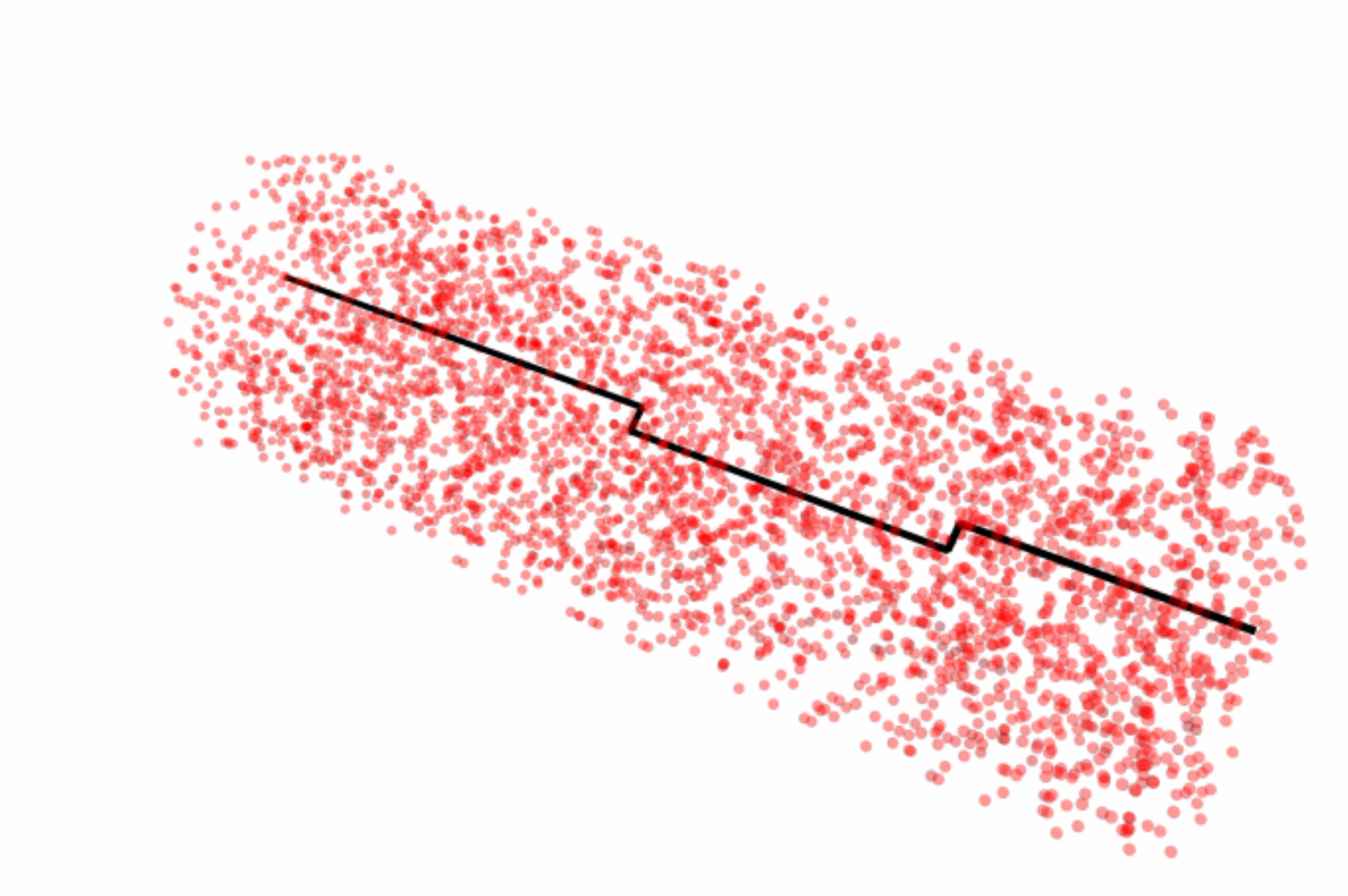}
        \caption{}
    \end{subfigure}
    \caption{Views of the solute prism containing a screw dislocation with a kink pair (a) along the dislocation direction and (b) general view. The images correspond to a simulation at 300 K, 200 MPa of applied stress, and 5\% Re concentration.\label{rojo}}
\end{figure*}

Our simulations involve periodic boundary conditions along the Burgers vector direction \cite{stukowski2015thermally}, while the other two dimensions are defined by the extent of the solute cloud around the dislocation. For convenience, we construct an hexagonal prism centered along the axis of the center-of-mass (COM) of the dislocation, with edges along the different glide planes of the $[111]$ zone. Then solute is randomly placed with the appropriate concentration on lattice sites contained within the prism prior to starting the simulations.
Because solute strain fields are short-ranged and solute diffusion occurs by way of discrete first-nearest-neighbor jumps, the size of the prism is not very important. Rather, the critical aspect here is to ensure that the prism is updated as the dislocation glides so that `fresh' material is generated ahead of the dislocation path. This is done to avoid periodic boundary artifacts resulting from prior dislocation-solute interactions. In this fashion, after every time step, the prism is shifted by the amount displaced by the dislocation COM, Re atoms belonging to the previous prism are kept, atoms outside the new prism are discarded, and regions of the prism without solute are populated randomly with the correct $c$. Figure \ref{rojo} shows images of the solute prism for a dislocation containing a kink-pair in W-5Re (at\%).

\section{Results}
\label{results}

The raw output of our method are dislocation velocities as a function of all the internal variables of the model: $T$, $c$, $\matr{\sigma}$, $L$, and MRSS plane. Here we study the variation of the flow stress with Re content, as the impact of the other variables on dislocation motion has been studied in depth in our previous work \cite{stukowski2015thermally}.

\begin{figure}[h]
	\centering
	\includegraphics[width=1.0\columnwidth,trim={2cm 1.5cm 0cm 15cm},clip]{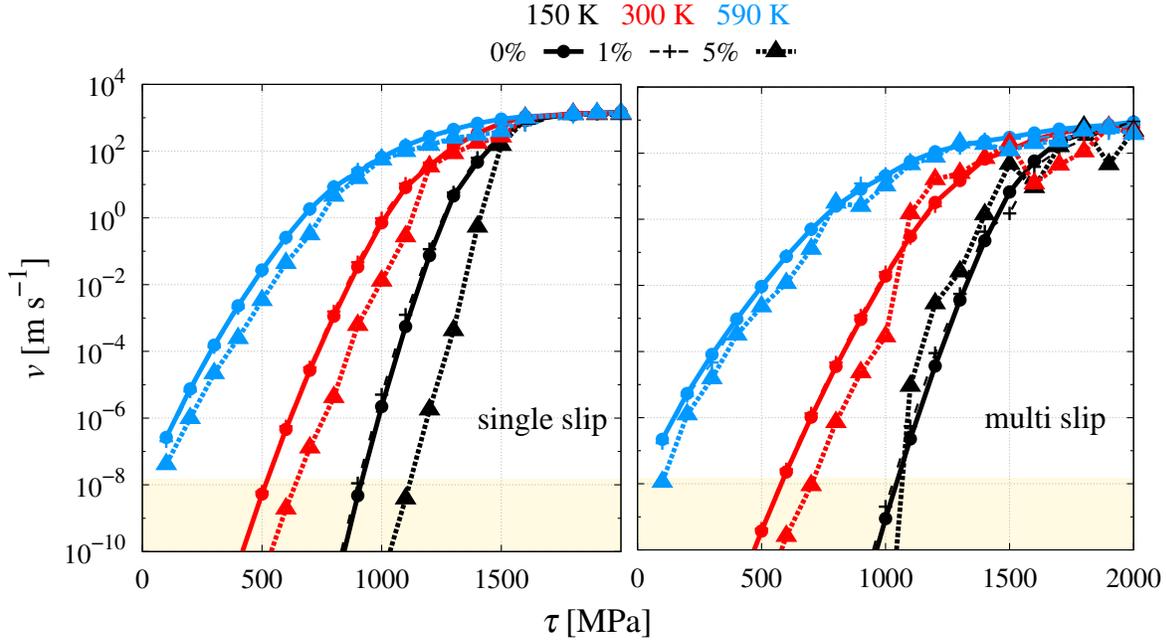}
	\caption{Dislocation velocity as a function of stress, temperature, and Re content. The data correspond to a screw dislocation segment $400b$ in length ($\sim$100 nm). Results for (nominally) single slip --MRSS plane is of the \{110\} family-- (left) and multislip --\{112\}-type MRSS plane-- (right) conditions are shown. The yellow shaded area marks the region where the dislocation velocity is not fast enough to activate slip, $v$$<$$v^{\ast}$ (see text).}
	\label{400b}
\end{figure}

Here we define the flow stress as the stress required to accelerate a screw dislocation to a velocity commensurate with the prescribed strain rate. This velocity can be calculated via Orowan's equation as:
$$v^{\ast}(\tau_f)=\frac{\dot\varepsilon}{\rho b}$$
where $\tau_f$ is the aforementioned flow stress, and $\rho$ is the dislocation density.  For the nominal conditions used by Stephens \cite{Stephens1970}, i.e.~a strain rate of $\dot\varepsilon=5.5\times10^{-4}$ s$^{-1}$, and a dislocation density $\rho\approx1.4\times10^{14}$ m$^{-2}$, we obtain a threshold dislocation velocity of $1.5\times10^{-8}$ m s$^{-1}$. By way of example, we first show screw dislocation velocities as a function of temperature, Re concentration, and MRSS plane in Figure \ref{400b}. For consistency with the experimental measurements, we consider only the case where $L=\rho^{-\nicefrac{1}{2}}\approx400b$ ($\approx$100 nm), and so we eliminate the dislocation length from consideration as a variable going forward. The figure shows results for 0, 5, and 20\% Re concentration at 150, 300, and 590 K. The yellow shaded area marks the region where the dislocation velocity is not fast enough to activate slip, i.e.~$v$$<$$v^{\ast}$, and thus from the intercepts of the velocity curves with the $v^{\ast}$ abscissa one can obtain the flow stress $\tau_f$ in each case. 
In other words, we obtain $\tau_f$ as the stress above which the dislocation velocity for each simulation triad $(L,T,c)$ is larger than $v^{\ast}$. 

It is helpful to consider the analytical analog of this flow stress for 0\% Re content to better understand its meaning. A closed-form expression for $\tau_f$ can be obtained assuming that the time employed by single kinks to sweep across the dislocation line, $\delta t_{\rm k}$ (cf.~eq.~\eref{tkp}), is much shorter than the time required to nucleate kink pairs,  $\delta t_{\rm k}=r_{\rm kp}^{-1}$ (cf.~eq.~\eref{kprate}). In such a case, the dislocation velocity can be approximated by \cite{cereceda2016unraveling}:
\begin{equation}
v\approx\frac{\nu_0h(L-w)}{b}\exp\left\{\frac{\Delta H_0}{kT}\left(1-\left(\frac{\tau_{\rm RSS}}{\sigma_P}\right)^p\right)^q\right\}
\end{equation}
where for simplicity we have replaced $\Theta(\matr{\sigma})$ in eq.~\eref{deltah} with the ratio $\tau_{\rm RSS}/\sigma_P$. Solving for $\tau=\tau_f$ when $v$=$v^\ast$ leads to:
\begin{equation}
\tau_f=
\sigma_P\left[
1-\left[
\frac{kT}{\Delta H_0}\log\left(\frac{\nu_0h(L-w)}{v^{\ast}b}\right)
\right]^{\frac{1}{q}}
\right]^{\frac{1}{p}}
\end{equation}
The above equation can be reduced to an expression of the type:
\begin{equation}
\tau_f=\sigma_P\left(1-(b_3T)^{\frac{1}{q}}\right)^{\frac{1}{p}}
\label{eqf}
\end{equation}
where the constant $b_3=\frac{k}{\Delta H_0}\log\left(\frac{\nu_0h(L-w)}{v^{\ast}b}\right)$ takes a value of $1.55\times10^{-3}$ K$^{-1}$ for $L=400b$ and the parameters given in Table \ref{tab-param}. 
\begin{figure}[h]
	\centering
	\includegraphics[width=1.0\columnwidth,]{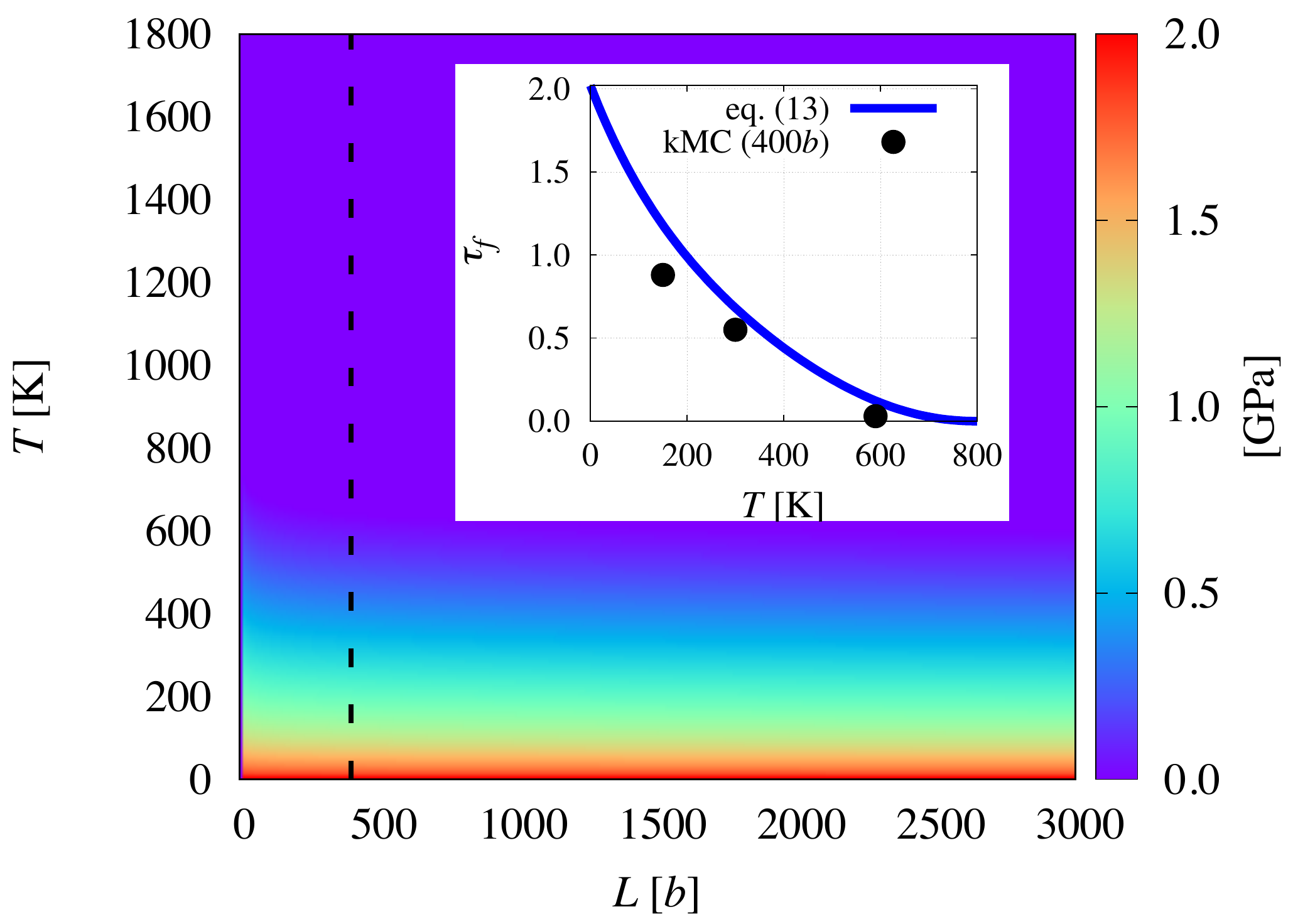}
	\caption{Contour map of the flow stress $\tau_f$ as a function of dislocation length $L$ and temperature $T$ according to eq.~\eref{eqf}. The inset shows $\tau_f$ at $L$=$400b$ (marked with a dashed line in the contour plot), as well as the three points obtained from the intercepts for 0\% Re in Fig.~\ref{400b}.}
	\label{contour}
\end{figure}
Equation \eref{eqf} is plotted as a contour map in Figure \ref{contour}. The inset to the figure shows the temperature variation of $\tau_f$ for $L$=$400b$ (marked with a dashed line in the main plot). More importantly, the three points obtained via kMC simulations in Fig.~\ref{400b} for 0\% Re are also plotted to confirm the consistency between the analytical and numerical approaches\footnote{The disagreement can be attributed to the absence of non-Schmid stresses in the analytical approach, and the stochastic variability of the kMC simulations.}. Next, we calculate $\tau_f$ as a function of $c$ and $T$ following the intercept technique discussed above.
\begin{figure}[h]
        \centering
	\includegraphics[width=0.65\columnwidth]{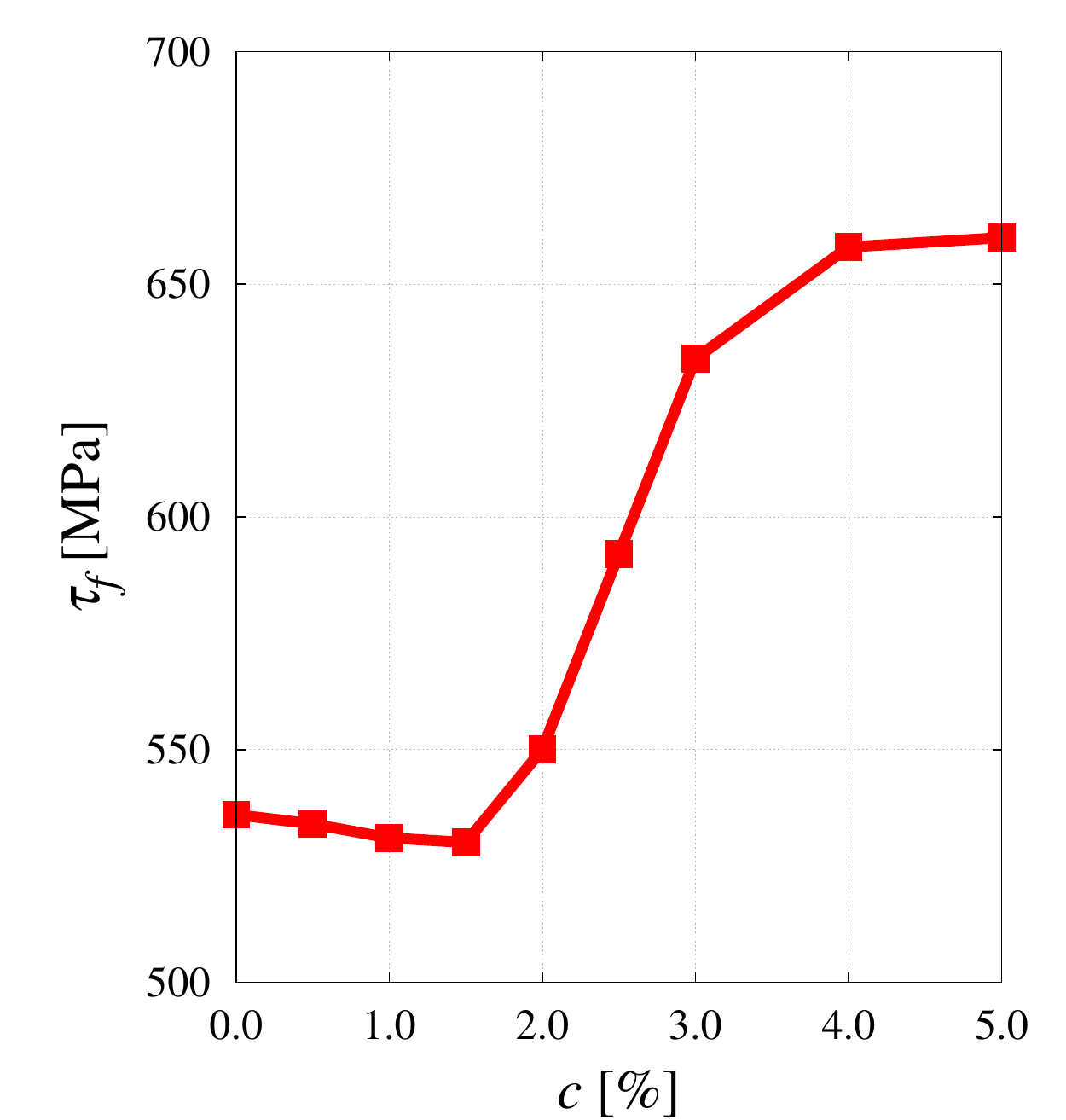}
	\caption{Flow stress variation with Re concentration at 300 K under \{110\} loading conditions.}
	\label{solo}
\end{figure}

\begin{figure}[h]
        \centering
	\includegraphics[width=0.65\columnwidth]{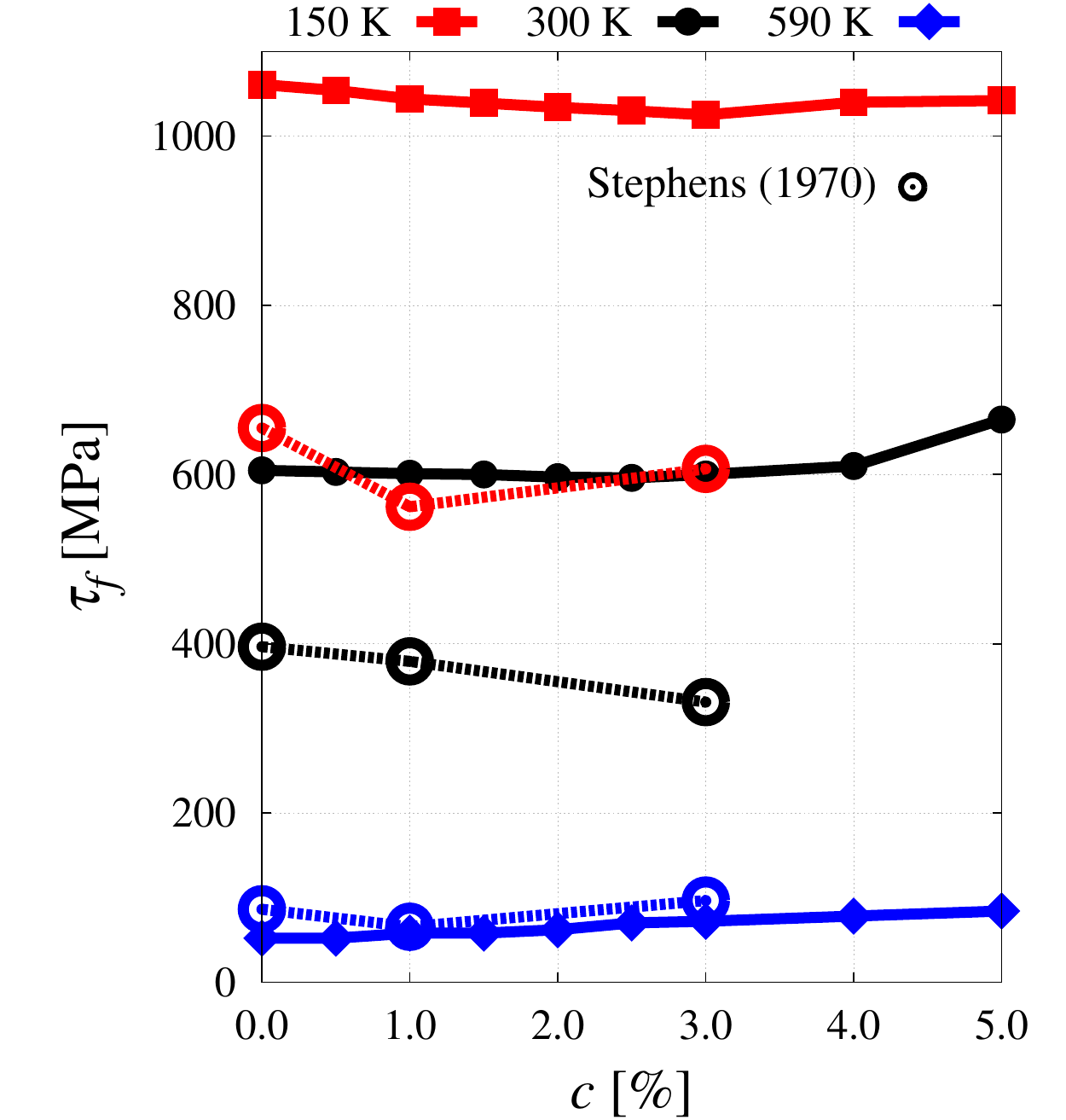}
	\caption{Dependence of the flow stress on Re concentration and temperature. The experimental measurements by Stephens are given for comparison\cite{Stephens1970}.}
	\label{tauf}
\end{figure}
The numerical results for $\tau_f$ as a function of Re content at 300 K under single slip conditions are shown in Figure \ref{solo}, with the minimum at $c$=1.5\%. The general dependence on temperature is shown in Figure \ref{tauf}. All curves show a region of softening at low solute concentrations, with a gradual transition to a hardening regime at higher values of $c$.
The experimental data by Stephens \cite{Stephens1970} are shown in the figure for comparison. As shown, the agreement with the measurements increases with temperature although it must be noted that the experimental data were obtained from full stress-strain curves, while our definition of $\tau_f$ is for a system with just one dislocation. 
\begin{figure}[h]
	\centering
	\includegraphics[width=0.75\columnwidth]{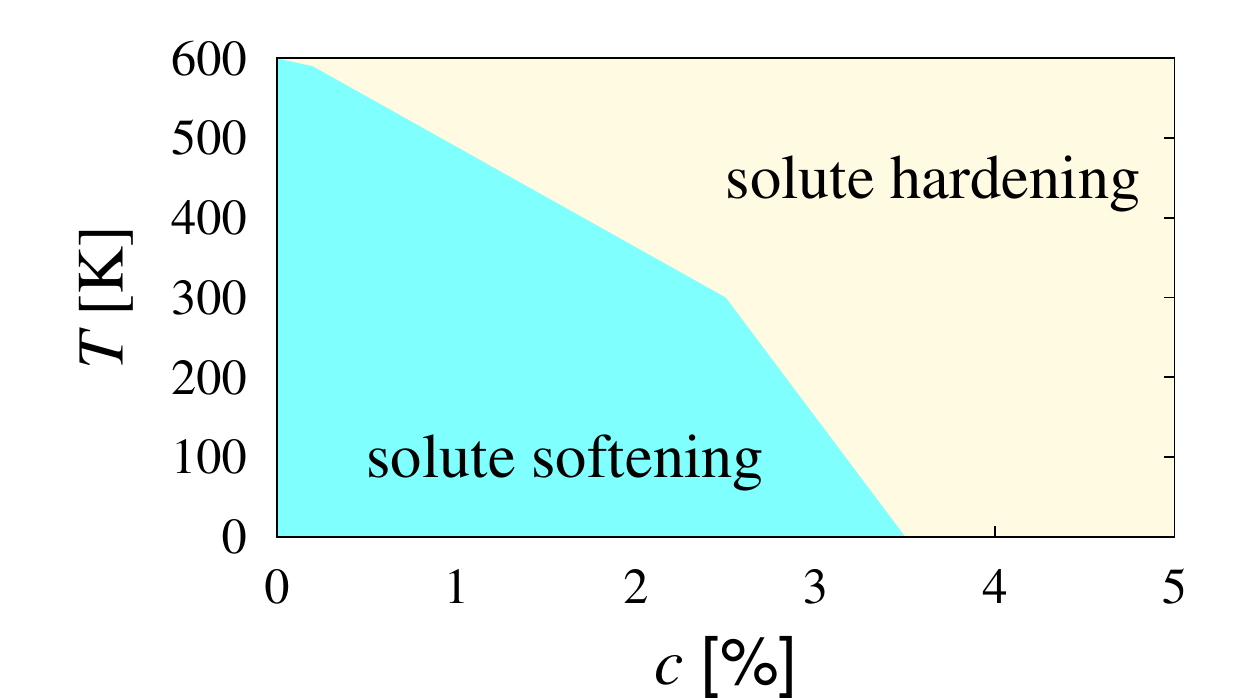}
	\caption{Solute softening/hardening map as a function of temperature and concentration under multislip conditions.}
	\label{map}
\end{figure}
The location of the minima at each temperature can be plotted into a phase diagram that delineates the separation of the softening and hardening regimes. This is seen in Figure \ref{map}, which can be considered a preliminary design map. We do not expect a strong dependence of the transition stress on dislocation line length in the laminar slip regime (only one kink-pair at a time on the dislocation), and so these results are in principle extrapolable to systems with higher dislocation densities such as W-Re in the cold-worked state.

\section{Discussion}
\label{disc}

The main purpose of this work is to assemble a computational model that incorporates the main physical mechanisms of substitutional solute-screw dislocation interactions in a bcc alloy. The motivation for the development of the model was to test the applicability limits of the softening effect of Re solute atoms on the Peierls stress of screw dislocations in W-Re, which is a general phenomenon known to occur in most bcc metals \cite{Trinkle1665}. 
This limit manifests itself as a minimum in the yield/flow stress-concentration curve, which is known to cause shifts in the DBTT compared to pure W.
Characterizing this transition from solid solution softening to solute hardening in substitutional bcc alloys has been the subject of much discussion in the literature \cite{RAVI1969547,RAVI1970623,Leslie1972,GIBALA19731143,PINK19801,LUO1991107,Okazaki1996}. It is generally well accepted that, before the minimum, the presumed softening of the Peierls stress with solute content aids kink-pair nucleation while having a negligible impact on kink-pair propagation. Above the minimum concentration, however, nucleation cannot compensate for the continued interference of solute atoms on kink motion. While there have been early semi-analytical models that build on this idea to predict the transition temperature and solute content, they have been mostly phenomenological due to a lack of understanding of the atomic-scale mechanisms involving dislocation cores and solute atoms. For example, Sato \etal~\cite{SATO1973753} were among the first to develop models to explain the transition in terms of the motion of a screw dislocation through a Peierls potential dotted with misfit strain centers representiing solute atoms. As well, there are more recent studies that also capitalize on these basic phenomenological premises \cite{KIRCHHEIM2012767,HU2017304}. Here, not unlike the work by Sato \etal, we develop a model based on the motion of a screw dislocation line through a solid solution. What separates our model from other approaches is the consideration of two of the most critical features of screw dislocation glide in bcc metals: thermally-activated kink-pair nucleation and non-Schmid effects. We have spent a great deal of time quantitatively characterizing these two \cite{stukowski2015thermally,po2016phenomenological,cereceda2016unraveling}, as well as the solute-dislocation interaction energetics for W-Re from first-principles calculations \cite{romaner2010effect,hossain2014stress}. This has opened up the possibility to perform \emph{predictive} simulations, which is what we report in this paper.

Our method has two restrictions that are worth noting. First, it is limited to a single dislocation, which ignores potential interactions with other dislocations and collective dislocation effects. However, there is now a growing consensus that the plastic response of most bcc metals and alloys can be characterized in terms of single-screw dislocation properties \cite{caillard2010_I,caillard2010_II,tang2014}. This is particularly true at low homologous temperatures, where the temperature dependence of the flow stress is particularly pronounced and edge dislocations play no discernible role on plastic flow. It is worth mentioning that experiments have shown that Re additions to tungsten promote the formation of edge dipoles at temperatures which are otherwise too low to hold any significant edge dislocation populations \cite{Stephens1970}, e.g.~300~and 150 K,  and increase dislocation density at all temperatures. While this may change the global balance of dislocation subpopulations, we believe that the softening-to-hardening transition is still controlled by local dislocation-solute interactions, well captured in this model.
Second, our framework admits only $\{110\}$ slip\footnote{That is, kink pairs can only form on $\{110\}$ planes, regardless of the MRSS plane.}, while it has been postulated that \{112\} slip may represent the primary slip plane at higher temperatures. While this has not been conclusively established \cite{seeger1981,Brunner_Wtemperature2010,marichal2013}, we have shown in prior works that the temperature dependence of the flow stress in single crystal W can be explained resorting solely to \{110\} slip \cite{cereceda2016unraveling}. Indeed, the results reported here show that the transition can be captured considering \{110\} glide only. It has been claimed that rhenium can increase the likelihood of slip on multiple \{110\} planes, even when the MRSS plane itself belongs to the \{110\} family. We have conducted analyses of the dislocation core trajectories as a function of concentration and have not found significant differences between pure W and W-Re\footnote{Instead, temperature is found to have a more pronounced impact on trajectory.}. In any case, we take our method to be strictly valid in the lower (homologous) temperature region of the flow stress curve, which is the regime considered here.

Finally, this work follows the line of research initiated by Cai and Bulatov \cite{cai1999kinetic,cai2001kinetic} and Deo and Srolovitz \cite{DEO308,DEO20051223}, where the foundations for the current model were laid. Extension of the current methodology to interstitial solutes and the study of dynamic strain aging is being currently pursued.

%
 
\section*{Acknowledgments}
The authors acknowledge support by the National Science Foundation under grant DMR-1611342, and the US Department of Energy's Office of Fusion Energy Sciences, project DE-SC0012774:0001. Computer time allocations at UCLA's IDRE Hoffman2 supercomputer are also acknowledged. Helpful discussions with A. Stukowski, V. Bulatov, and D. Rodney are gratefully acknowledged.

\section*{References}
\bibliographystyle{iopart-num.bst}
\providecommand{\newblock}{}

\end{document}